\documentclass[sigconf]{acmart}

\copyrightyear{2026}
\acmYear{2026}
\setcopyright{cc}
\setcctype{by}
\acmConference[UIST '26]{The 39th Annual ACM Symposium on User Interface Software and Technology}{November 02--05, 2026}{Detroit, MI, USA}
\acmBooktitle{The 39th Annual ACM Symposium on User Interface Software and Technology (UIST '26), November 02--05, 2026, Detroit, MI, USA}
\acmDOI{10.1145/3830398.3830629}
\acmISBN{979-8-4007-2856-3/2026/11}
\AtBeginDocument{%
  }

\newcommand{\appname}{\textsc{Zoro}}
\usepackage{xcolor}
\usepackage{varwidth}
\usepackage{paracol}
\usepackage{soul}
\usepackage{listings}
\usepackage{xcolor}
\usepackage{makecell} 

\usepackage{tikz}
\usetikzlibrary{shapes.geometric}

\DeclareRobustCommand{\blueletterbadge}[1]{%
  \tikz[baseline=(char.base)]{
    \node[
      circle,
      fill={rgb,255:red,0;green,132;blue,255},
      text=white,
      inner sep=0pt,
      minimum size=2.5ex,
      font=\bfseries\footnotesize
    ] (char) {#1};
  }%
}

\DeclareRobustCommand{\grayletterbadge}[1]{%
  \tikz[baseline=(char.base)]{
    \node[
      circle,
      fill={rgb,255:red,255;green,0;blue,0},
      text=white,
      inner sep=0pt,
      minimum size=2.5ex,
      font=\bfseries\footnotesize
    ] (char) {#1};
  }%
}

\DeclareRobustCommand{\grayletterbadge}[1]{%
  \tikz[baseline=(char.base)]{
    \node[
      circle,
      fill={rgb,255:red,255;green,178;blue,0},
      text=white,
      inner sep=0pt,
      minimum size=2.5ex,
      font=\bfseries\footnotesize
    ] (char) {#1};
  }%
}

\usepackage{xcolor}
\usepackage{framed}
\usepackage{listings}

\newcommand{\RuleBox}[2]{%
  \par\medskip
  \noindent
  \fcolorbox{gray!60}{gray!10}{%
    \parbox{\dimexpr\linewidth-2\fboxsep-2\fboxrule\relax}{%
      \scriptsize
      \textbf{#1}\par
      \textit{Verification:} #2%
    }%
  }%
  \par\medskip
}

\newenvironment{GrayPanel}{%
  \MakeFramed{\advance\hsize-\width\FrameRestore}%
  \small\sffamily
}{%
  \endMakeFramed
}

\definecolor{diffgreen}{HTML}{D1FFD1}
\sethlcolor{diffgreen}

\definecolor{codegray}{rgb}{0.95,0.95,0.95}
\definecolor{rulebg}{RGB}{245,245,245} 
\newcommand{\change}[1]{\textcolor{black}{#1}}

\begin{document}

\title{\appname: \change{Active Controls to Support Rule Following in Vibe Coding}}

\author{Jenny Ma, Sitong Wang, Joshua H. Kung, Lydia B. Chilton}
\affiliation{%
  \institution{Columbia University}
  \city{New York}
  \state{New York}
  \country{USA}\\
  \href{mailto:jenny.ma@columbia.edu, sitong@cs.columbia.edu, jhk2214@columbia.edu, chilton@cs.columbia.edu}{\texttt{\{jenny.ma, sitong, chilton\}@cs.columbia.edu}, \texttt{jhk2214@columbia.edu}}
}








\renewcommand{\shortauthors}{Ma et al.}

\begin{abstract}
  Rules files (e.g., \texttt{AGENTS.md}, \texttt{CLAUDE.md}) are the primary mechanism for human-agent alignment when developers vibe code. However, they remain passive: it is not immediately apparent when rules are being used or followed, or how to improve them. To transform rules from passive text into active controls, we introduce \appname, an interactive interface that integrates directly with a coding agent and anchors rules to every step of the coding process. After an agent generates an initial plan, \appname\ \textit{enriches the plan} with rules, \textit{enforces the rules} during implementation by requiring the agent to prove that each rule was followed, and allows users to provide in-situ feedback when they are unsatisfied with a rule application to \change{\textit{evolve the scope of the rules}}. A technical evaluation shows that coding agents follow rules more with \appname\ than without. 
  A user study demonstrates a change in people's behavior and cognitive strategies when rules are at the forefront of vibe coding. 
  We discuss how making rules active in agentic systems unlocks broader opportunities for human-agent alignment in coding settings and beyond.
\end{abstract}




\begin{teaserfigure}  
    \centering
    \includegraphics[width=\textwidth]{figures/system_eee.png}
    \caption{\textit{Enrich-Enforce-Evolve} is a framework for making rules active. \change{In \textit{Enrich}, multiple rules, including \textit{Rule 7}, are retrieved and used to reshape the plan (e.g. substeps were added to \textit{Step 1}) before the agent executes the plan. In \textit{Enforce}, we show the agent executing \textit{Step 1b}. The agent must both generate code \textit{and} prove that \textit{Rule 7} is being followed through code evidence and unit tests. A user adds \textit{in-situ notes} to \textit{Rule 7}'s rule application. After the full plan is executed, during \textit{Evolve}, we aggregate all \textit{in-situ notes} to refine and further specify the rules so that they are grounded in real examples.}}
    \label{fig:system_eee}
\end{teaserfigure}

%
\maketitle

\section{Introduction}
Developers no longer have to write code line-by-line. Instead, they steer agents through iterative prompting, a process known as vibe coding \cite{Karpathy2025VibeCoding, GoogleVibeCoding2025}. But prompting is unreliable;  agents often ignore instructions, and developers are left repeating themselves session after session.
Rules files (e.g., \texttt{AGENTS.md} \cite{AgentsMD2024}, \texttt{CLAUDE.md} \cite{AnthropicMemory2025} or \texttt{.md} files in custom rule directories \cite{Cline2025Rules}) address this by encoding persistent intent ("always use brand colors in the UI", "never access database directly from frontend", etc.) -- giving developers a way to embed coding conventions, architectural constraints, and workflow preferences across sessions. Widely adopted across industry tools 
\cite{github2026copilotinstructions, cursor2026rules, anthropic2026memory, 
openai2026agentsmd}, they have emerged as the primary mechanism for human-agent alignment in vibe coding, letting developers personalize and guide agent behavior without micromanaging each step. 

Despite the importance of rules, they remain passive text. 
Rules are loaded once at the start of a coding session then gradually lose relevance, so their influence deteriorates throughout the session \cite{anthropic2026memory,github2026copilotinstructions,liu2024lostmiddle,laban2026lostconversation}. 
Additionally, developers can't easily tell if an agent actually followed rules at 
all~\cite{shankar2024validators, epperson2025agdebugger}. 
Finally, developers tend to append rules reactively rather than regularly refine them, leading to outdated or repetitive rules.
Thus, they remain in a defensive role – constantly monitoring outputs, correcting mistakes, and reasserting rules that were already written down ~\cite{barke2023grounded,Wang_2024}. 
Given the importance of rules, how can we surface them so that they become a focal lever — something developers can actively see, engage with, and rely on?

We propose transforming rules from passive text into active controls.
Based on findings from our developer study, 
we introduce a framework called \textit{Enrich-Enforce-Evolve} to push rules to the forefront of the developer's attention (see Figure \ref{fig:system_eee}).
The first principle is to \textit{enrich} the plan: after a coding agent generates an implementation plan, relevant rules are fetched to rewrite the plan and anchor each step of the plan to rules. 
As the agent executes the plan, rules are \textit{enforced}: the agent must follow each rule associated with the current step by providing code evidence and unit tests.
Developers review this evidence and add in-situ notes where rules fall short of their standards. Finally, rules \textit{evolve}. At the end of the session, the in-situ notes, each anchored to a concrete example of a rule application, are aggregated and used to sharpen rules and capture preferences that emerged in practice. 
\change{Ultimately, \textit{Enforce} is the fundamental principle that transforms rules from static guidance into active, verifiable controls; \textit{Enrich} and \textit{Evolve} provide the surrounding workflow to make enforcement feasible in practice.}

 %


We introduce \appname, a plugin layered on top of a coding agents
that instantiates \textit{Enrich-Enforce-Evolve} and anchors rules to every phase of vibe coding.  
We focus on vibe coding--where developers steer agents entirely through natural language rather than writing code directly-- rather than AI-augmented software development, as it represents a rapidly growing practice with a large user base and a well-constrained starting point~\cite{Karpathy2025VibeCoding, GoogleVibeCoding2025}.
Rather than rules sitting in a markdown file hoping to be noticed, \appname\ surfaces them as living, visible objects: users see which rules apply at each step, see them get enforced in real time, and refine them based on what they observe.
In \appname, developers see which rules apply at each step, watch them get enforced in real time, and refine them based on what they observe. 
Users interact with Zoro via a front-end interface to create, modify, and review rules, while simultaneously using the native coding agent chat interface to direct the agent to code. 
A technical evaluation demonstrates that \appname\ increases rule 
following \change{from 51\% to 80\%}.
Our user study reveals that surfacing rules as active controls changes how developers approach vibe coding; specifically we notice a shift from prompt engineering to rule engineering as a method of controlling agents.
We conclude with a discussion on how making rules active unlocks broader opportunities for human-agent alignment in coding settings and beyond.
Our contributions are as follows: 
\begin{itemize}
     \item \textbf{Enrich-Enforce-Evolve}: \change{a framework to transform passive rules into active controls. We first \textit{enrich the plan} in order to \textit{enforce the rules}, then \textit{evolve the scope of the rules} for future enforcement. \textit{Enrich} and \textit{Evolve} work in service of \textit{Enforce.}}
    \item \textbf{\appname}: \change{An enforcement layer integrated with a coding agent that instantiates \textit{Enrich-Enforce-Evolve}.}
    \item \textbf{Technical Evaluation}: A controlled evaluation of 36 vibe coding sessions demonstrating that \appname\ increases rule following for coding agents.
    \item \textbf{User Study}: A qualitative study (N=12) examining how vibe coding is transformed when rules are made active. 
\end{itemize}
\section{Related Work}
\subsection{Lack of Transparency in Vibe Coding}
While AI coding tools are powerful and widely valued, prior work has found that developers 
frequently reject generated code because it fails to meet user's 
requirements and is difficult to steer~\cite{liang2024large}. Programmers must actively 
ground, inspect, and repair AI-generated code within its context to ensure alignment with 
their intent~\cite{barke2023grounded}, and this challenge intensifies during iterative 
exploration~\cite{zamfirescu2025beyond} or when working with complex codebases~\cite{yang2025trumanbench}.

Beyond general usability concerns, recent work frames trust and validation as first-order challenges in AI-assisted programming. Wang et al. show that developers struggle to form appropriate trust in AI-powered code generation, especially around expectation-setting, configuration, and suggestion validation~\cite{Wang_2024}. Ferdowsi et al. find that runtime validation of AI-generated code fails because execution-based inspection is too disruptive, and show that continuously surfacing runtime values during editing reduces this problem~\cite{ferdowski_2024}. Together, this work suggests that reliable AI-assisted programming depends not only on better generation, but on interfaces that help developers inspect and verify behavior. \appname\ addresses this gap by explicitly surfacing developer rules, enabling users to specify, verify, and refine them in context.

\subsection{Context Engineering and Workflow Alignment}

Recent work has improved alignment by helping AI systems build better context over the course of interaction. 
Most of this work focuses on two capabilities: \textit{enriching} context with more relevant user information, and \textit{evolving} that context as user intent changes. 
For example, prior systems infer preferences from interaction histories~\cite{shaikh2025creating}, curate personal context for planning~\cite{zhang2024jumpstarter}, and iteratively specify preferences through in-situ notes and feedback~\cite{vaithilingam2025semantic, shankar2025steering, petridis2024constitutionmaker}. \change{Like prior work, \appname\ \textit{enriches} the plan with rule context and \textit{evolves} individual rule specifications to keep them from drifting over time.}
These approaches make system context richer and more adaptive, but they stop short of ensuring that the resulting context is actually followed and \textit{enforced} during execution. 

This gap is especially important in workflow-based settings, where alignment must persist throughout a multi-step process rather than appear only in an initial prompt or plan. 
In coding agents like Cline ~\cite{Cline2024}, existing enforcement is typically ad hoc rather than systematically integrated into the workflow itself. 
Recent systems have started treating alignment as something that must be maintained across planning and execution because intentions can drift once execution begins ~\cite{he2025plan, feng2024cocoa}.
Other systems further suggest that alignment benefits when agents create structured moments for oversight during execution~\cite{horvitz1999mixedinitiative, chen2025needhelp, pu2025assistancedisruption, epperson2025agdebugger, peng2025morae, ma2025agentdynexnudgingmechanicsdynamics}. 
For instance, Need Help? surfaces proactive assistance before users explicitly ask for it~\cite{chen2025needhelp}, while Morae pauses UI agents at decision points so users can intervene on meaningful choices~\cite{peng2025morae}. 
We build on this line of work by treating developer rules as process-level alignment constraints: rules are not just context to read, but checkpoints that are actively \textit{enforced} and updated throughout execution.

\subsection{Anchoring Rules in Execution}
Developers increasingly externalize intent into rule artifacts such as \texttt{AGENTS.md} files and agent scaffolding specifications (e.g., harness engineering~\cite{lopopolo2026harness}). Yet, 
when served as static context, they do not reliably improve task success and can increase exploration cost~\cite{gloaguen2026evaluating}, largely due to context rot \cite{liu2024lostmiddle,laban2026lostconversation}. 
Prior work suggests that intent must be anchored to agent actions at each step.
Existing systems build on this; DynEx grounds high-level intent into structured intermediate designs~\cite{ma2025dynex}. Another study shows users benefit from visible intermediate assumptions and 
step-level verification~\cite{kazemitabaar2024}. 

A broader line of systems ties model behavior to concrete evidence at the point of action. 
Within programming, Ivie anchors lightweight explanations directly to just-generated code so developers can interpret outputs in relation to the code itself~\cite{yan2024ivie}, and Trailblazer answers developer questions with annotated, replayable agent-discovered program traces~\cite{yan2025trailblazer}. 
CodeA11y connects accessibility intent to concrete coding actions by pairing generation with accessibility-focused guidance in context~\cite{mowar2025codea11y}. 
Beyond programming, InkSync ties model-suggested edits to executable document operations~\cite{laban2024inksync}, while HaLLMark grounds AI writing assistance in interaction history, exposing where model contributions came from and how they entered the workflow~\cite{hoque2024hallmark}. 
Collectively, this work suggests that human-AI systems are more usable when model behavior is tied to concrete intermediate evidence~\cite{liao2024transparencyagenda,horvitz1999mixedinitiative}. 
Our work builds on this by anchoring rules to execution steps, showing the rule enforcement, and refining rules based on grounded evidence.





\section{Design Goals}

To better understand the challenges developers face when vibe coding with rules, we conducted semi-structured interviews with 10 expert programmers (5-10 years of coding experience) \change{ recruited via snowball sampling. Participants regularly vibe code with rules files and with their primary coding agent spanning different coding agents: Claude Code, Cursor, Codex, and Cline.} Additionally, we observed three programmers vibe code with their own rules; each session lasted 2-3 hours. \change{We analyzed the interview transcripts using an iterative thematic analysis process to identify recurring challenges and motivate the design of \appname\ rather than to provide generalizable claims about developer behavior.}

We found that developers largely appreciate rules files and believe they improve agent performance. Participants generally knew what rules they wanted; they either had rules files written from personal experience or adapted from templates shared online. P3 noted: “[Rules files] definitely work…still not perfect. . .[but] the better context you give AI, the better it’s going to perform.” \textbf{However, there is little visibility into how effective individual rules actually are}
\textit{(Challenge 1: Rule Invisibility)}. Developers largely have a general sense that rules helps, rather than understanding which rules work and why. P3 said, “it’s difficult to know when you have a good [rule] versus a bad [rule], it's very hard to find the balance.” 

\textbf{Participants also reported issues of rules not being followed} \textit{(Challenge 2: Skipping Rules)}. There is the problem of context rot – rules are loaded initially but not continuously retrieved – leading agents to gradually stop following them. P3 described his personal experience: “a very common rule it messes up is not using the ORM (Object-Relational Mapping). Even though it is specified in multiple places [in the rules file], it still tries to make a direct database call.” \textbf{Developers also admitted to not properly maintaining their rules files and just adding to them without refining existing rules, resulting in increasingly messy and inconsistent rules} \textit{(Challenge 3: Ruleset Decay)}. P1 noted that when “[the \texttt{AGENTS.md}] just becomes too bloated… [the agent starts] applying things where it shouldn’t.”

We outline \change{three} key design goals (DGs) based on the above challenges. Since developers already have rules, our design goals focus not on rule creation, but on making existing rules visible, enforceable, and improvable throughout the vibe coding workflow: 


\begin{itemize}
    \item \textbf{DG1: Make Relevant Rules Explicit.}  Plans are generated without explicit references to rules, and developers have little visibility into which rules influence execution. 
    Systems should therefore surface relevant rules and integrate them directly with the plan and make their influence explicit. 
    \item \textbf{DG2: Guarantee that Rules are Followed.} Developers need to trust that important rules are actually followed. Systems should enforce rules during code implementation and prevent progress when rules are violated. They should also provide evidence linking rules to specific code, allowing developers to check whether the rule had the intended effect. 
    \item \textbf{DG3: Enable Iterative Refinement.} Rules are rarely maintained, largely because developers lack a clear way to evaluate their effectiveness. Instead of “hoping for the best,” developers should be able to see when rules are not to their standard and immediately refine them.
\end{itemize}

\section{\appname}


Based on our design goals, we present \appname 
\footnote{\url{https://github.com/jennygzma/zoro}},
a system that augments vibe coding workflows by turning rules into active controls. \appname\ plugs into the user's coding agent of choice (Codex \cite{OpenAICodex2024}, Claude Code \cite{Anthropic2025ClaudeCode}, Cline \cite{Cline2024}, or Cursor \cite{Cursor2024}) and anchors rules to every step of the coding process. Users interact with \appname's front-end interface to review rules, while simultaneously directing their coding agent to write code. 
At the heart of \appname\ is \textit{Enrich-Enforce-Evolve}, which maps directly onto phases of vibe coding (planning, code implementation, and review):
\begin{figure*}[t]  
    \centering
    \includegraphics[width=\textwidth]{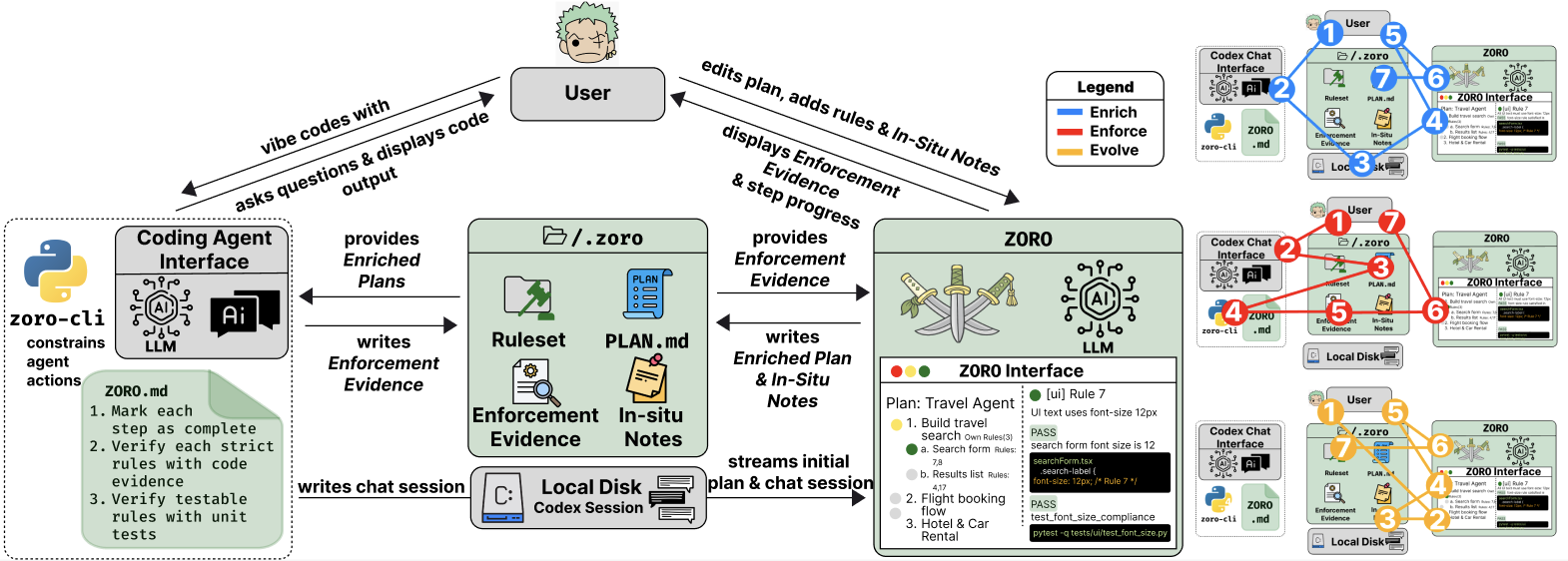}
    \caption{\appname\ Ecosystem. \change{\appname\ consists of two parts: 1) a python package called \texttt{zoro-cli} to constrain the agents actions during enforcement and 2) a rule visualization interface for the user to interact with. The \texttt{.zoro} folder in the target repository acts as a shared communication layer between \appname's interface and the coding agent. The left panel illustrates the overall system architecture, while the right panel visualizes the interaction and information flow for the Enrich (blue), Enforce (red), and Evolve (orange). The numbers trace how plans, enforcement evidence, and notes move between the interface, the \texttt{.zoro} repository, and the coding agent. }}
    \label{fig:system}
\end{figure*}
\begin{itemize}
    \item \textit{Enrich the Plan}. When the coding agent generates a plan, \appname\ reshapes the plan so that each step is anchored in rules, and each step explicitly cites relevant rules \textit{(DG1)}. \change{Enrichment builds on prior work that richer task context improves agent alignment~\cite{zhang2024jumpstarter}; here it also establishes the rules required for later enforcement.}
    \item \textit{Enforce the Rules}. As the agent writes code to execute the plan, \appname\ constrains the agent’s actions \change{with a CLI enforcement layer. Before advancing to the next step, the agent must invoke CLI commands that provide evidence demonstrating compliance with each applicable rule.} \textit{(DG2)}.
    \change{Although evidence-based self-verification is well established \cite{veriplan}, our contribution is this CLI enforcement layer that makes verification a required part of the coding workflow. }
    \item \textit{Evolve \change{the Scope of the Rules}}. Users review enforcement evidence to confirm their rules were followed and add in-situ notes if they dislike how a rule was applied. In-situ notes are aggregated and fed to an LLM to refine the rules scope and close the "rule-loop" \textit{(DG3)}. \change{Refining intent via in-situ notes builds on prior work \cite{shankar2025steering}; in \appname\, it ensures that future enforcement uses rules that reflect developer preferences.} 
   
\end{itemize}

\appname\ has two entry points into the developer's workflow, shown in Figure~\ref{fig:system}. The first is a visual interface,
where the user monitors execution, reviews enforcement evidence, and refines rules. 
The second is \texttt{zoro-cli}, a Python package installed at the repository root that plugs \appname\ directly into the coding agent's environment — the user continues vibe coding as normal, while \appname\ reads what the agent produces. \texttt{zoro-cli} ensures the coding agent enforces each rule, and surfaces everything for the user to verify.

To make this possible, \texttt{zoro-cli} installs two items in the project repository: 1) an instruction file called \texttt{ZORO.md} that tells the agent it must call \texttt{zoro~update-step} to mark progress and \texttt{zoro~prove-rule} to submit evidence for each rule before it can advance to the next step, and 2) a shared directory, \texttt{.zoro}, that acts as the communication channel between the agent and \appname. \appname\ writes the enriched plan there, the agent reads it and writes back its progress and evidence. Because all communication flows through this directory, \appname\ is fully agent-agnostic: any coding agent that can read files and run CLI commands can integrate with the system. Additional package and framework details are in Appendix \ref{app:system_libraryandframeworks}.

In this section, we provide a usage scenario to walk through \appname’s interface and implementation details.
Consider Johnny, an engineer at a mid-sized company building \textsc{Logpose}, an internal tool for HR teams to log employee sentiment and surface patterns across the organization. Because \textsc{Logpose} handles sensitive employee data, Johnny must maintain strict API boundaries, careful schema migrations, and a coherent design system. He uses Codex to vibe code, and over time has built up an \texttt{AGENTS.md} file with rules like always enforce API boundaries so the frontend never reaches directly into backend logic, reuse shared UI components instead of importing from external libraries, 
and always backfill existing data on schema changes. These rules reflect hard-won lessons from past sessions where things went wrong. But Johnny cannot tell when Codex actually follows the rules, and he has no way of knowing whether they are well-specified enough to be useful. To anchor rules in his development workflow, he decides to use \appname. \footnote{See Appendix \ref{app:system_setupandruleimport} and \ref{app:system_rulemanagement} for setup and rule creation details.}


 \begin{figure*}[t]  
    \centering
    \includegraphics[width=\textwidth]{figures/system_interface.png}
    \caption{\appname's Visualization Interface. Users can interact with the plan by adding or deleting rules from each step of the plan. As the agent codes, users can track its progress, audit enforcement evidence, and add in-situ notes. (More details on the \textit{Supervision}, \textit{Rule Learning}, \textit{Rule Review}, and \textit{Rule Management} tabs can be found in Appendix \ref{app:system_monitoring}, \ref{app:system_rulereview} , and \ref{app:system_rulemanagement}).}
    \label{fig:system_interface}
\end{figure*}

\subsection{Enriching the Plan}
Like any other normal Codex session, Johnny types his task directly into the Codex interface. This full-stack feature spans backend data modeling, LLM-based categorization, and front-end components:

\noindent
\fcolorbox{gray!60}{white}{%
\parbox{\dimexpr\linewidth-2\fboxsep-2\fboxrule\relax}{%
\scriptsize\fontfamily{phv}\selectfont
help me build a feature that auto-organizes employee log entries by sentiment category. managers should be able to preview the suggested groupings, drag and drop entries to adjust them, and apply the final organization
}}

Codex generates an initial development plan (see Appendix \ref{app:system_initialplan}). The plan is technically reasonable, but it doesn't reference any of Johnny's rules; without them, there's a danger that his rules won't be followed.
To view the plan in \appname, Johnny opens the Visualization tab. \appname\ has detected that Codex created a plan and enriched it. Johnny sees an interactive plan outline (Figure~\ref{fig:system_interface}\blueletterbadge{A}) — each step and substep is expandable, and rules appear embedded directly in the step it governs (Figure~\ref{fig:system_interface}\blueletterbadge{B}). Where Codex's plan simply says ``Add Category model to the backend schema,'' \appname\ has introduced a dedicated substep to handle the migration and attached two rules to it. 
The rules are now woven into the plan at the exact moment they need to be followed:

\noindent
\fcolorbox{gray!50}{white}{%
\begin{minipage}{\dimexpr\linewidth-2\fboxsep-2\fboxrule\relax}
\scriptsize\fontfamily{phv}\selectfont
\setlength{\parskip}{0pt}

\begin{center}
\textbf{Original Agent Plan}
\end{center}

\textbf{Step 1: Category System Foundation}\par
Step 1.1: Add \texttt{Category} model to backend schema (default category color: grey); Add \texttt{category\_id} field to \texttt{LogEntry} records\par
Step 1.2: Create a \texttt{Category} service in \texttt{api/categories.py} to support CRUD operations\par
Step 1.3: Implement a category sidebar view and integrate it with the main log entry interface\par
\textbf{Step 2: AI Auto-Organization...}\par
\textbf{Step 3: Preview Dialog and Frontend...}\par

\medskip
\hrule
\medskip

\begin{center}
\textbf{Zoro-Enriched Plan}
\end{center}

\textbf{Step 1: Category System Foundation}\par
Step 1.1: Add \texttt{Category} model to backend schema \hl{(default to user-specified color)}\par
- \hl{Ask the user to choose the default category icon color}\par
- Add \texttt{category\_id} to \texttt{LogEntry} records\par
- \hl{Backfill existing data safely}\par
\textbf{\hl{RULE A: Always ask the user before introducing a new color}}\par
\textbf{\hl{RULE B: Ensure all schema changes properly backfill or migrate existing data}}\par
Step 1.2: Create a \texttt{Category} service in \texttt{api/categories.py}\par
\textbf{\hl{RULE C: Follow existing repository patterns for consistency}}\par
\hl{Step 1.3: Define frontend types and register CRUD endpoints}\par
- \hl{Create \texttt{category} interface in \texttt{frontend/src/types}}\par
- \hl{Register endpoints in \texttt{frontend/src/services/api.ts}}\par
\textbf{\hl{RULE D: API methods must return typed domain objects}}\par
\textbf{\hl{RULE E: Keep frontend types and backend models synchronized}}\par
\hl{Step 1.4: Implement category sidebar and main log entry interface}\par
\textbf{\hl{RULE F: Handle deletions inline (avoid browser confirm dialogs)}}\par
\textbf{\hl{RULE G: Use shared design system components instead of MUI}}\par
\textbf{Step 2: AI Auto-Organization...}\par
\textbf{Step 3: Preview Dialog and Frontend...}\par

\end{minipage}%
}

\vspace{0.5em}

Before handing the plan to Codex, Johnny can edit step descriptions, reorder substeps, or insert new ones  (Figure~\ref{fig:system_interface}\blueletterbadge{C}). He also sets enforcement levels for each rule: \textit{non-strict}, where the rule surfaces as guidance but doesn't require proof; \textit{strict}, where Codex must provide explicit evidence before advancing; and \textit{testable}, where Codex must additionally run executable unit tests for each rule (Figure~\ref{fig:system_interface}\blueletterbadge{D}). \texttt{Rule A} is \textit{strict} — he does not want Codex quietly choosing a color. He marks \texttt{Rule B} as \textit{strict} and \textit{testable} — he has seen data inconsistencies compound before and wants proof the migration ran. \texttt{Rule C} is \textit{non-strict} — he wants the agent to be reminded of this rule but doesn't need hard evidence. With enforcement levels set, the plan is ready to be implemented.

\subsubsection{Implementation}
\appname\ detects that Codex has generated a plan by continuously polling Codex's local chat log, which Codex writes to disk (Figure \ref{fig:system} \blueletterbadge{2}\blueletterbadge{3}). 
Once the plan is detected, \appname\ sends it alongside Johnny's full ruleset to GPT-5.3, which cross-references the two and produces the enriched plan: decomposing steps, attaching relevant rules to each, and flagging appropriate enforcement levels (Figure \ref{fig:system} \blueletterbadge{4}). Rules are structured as JSON objects with metadata including enforcement level, category, confidence score, decay score, and contextual scope — this metadata gives the LLM rich context for deciding which rules are relevant to which steps. \appname\ writes the enriched plan to \texttt{logpose/.zoro} and renders it on the interface for the user to interact with (Figure \ref{fig:system} \blueletterbadge{5}\blueletterbadge{6}\blueletterbadge{7}).
\subsection{Enforcing the Rules}

Seeing that \appname\ has enriched the plan, Johnny types: "execute the plan!” on the Codex interface. 
On \appname's Visualization tab, Johnny watches \textit{Step 1.1} flip to a yellow ``in progress'' indicator  (Figure~\ref{fig:system_interface}\grayletterbadge{E}). Codex begins generating code for the \texttt{Category} model. Because \texttt{Rule A} is \textit{strict}, Codex pauses mid-execution and asks Johnny whether gray is an acceptable default color. Johnny chooses green. This is a rule that might normally go unnoticed, but it gets enforced because it is embedded into \appname's plan. Codex logs the evidence using \texttt{zoro prove-rule} and generates code for the rest of the step, then attempts to advance to \textit{Step 1.2}. The CLI stops it:


\vspace{0.5em}
\noindent
\fcolorbox{gray!60}{white}{%
\parbox{\dimexpr\linewidth-2\fboxsep-2\fboxrule\relax}{%
\sffamily\scriptsize
{Error:} Cannot mark step-1.1 as complete. \\
Unverified rules: \\
{rule-b} (Ensure all schema changes properly backfill or migrate existing data) \\
Please verify all strict rules using \texttt{zoro prove-rule} before completing the step.}}
\vspace{0.5em}

Codex is forced back to handle the migration. It generates a backfill script for all existing \texttt{LogEntry} records, runs a unit test to verify it, and submits proof. On the interface, Johnny watches the enforcement evidence roll in for \textit{Step 1.1} in real time. 
For both rules, the enforcement panel surfaces a plain-language summary of what was done, the relevant code artifact (Figure~\ref{fig:system_interface}\grayletterbadge{F}), and — for testable rules — the unit test (Figure~\ref{fig:system_interface} \grayletterbadge{G}, full evidence can be found in Appendix \ref{app:system_enforcementevidence}).
With both rules satisfied, Codex marks \textit{Step 1.1} complete. 
As the session continues and enforcement evidence accumulates, he has an audit trail grounded in actual code and tests.

\subsubsection{Implementation}
The plan is executed as follows: Codex reads the plan from \texttt{logpose/.zoro} and begins working through it step by step (Figure \ref{fig:system} \grayletterbadge{2}\grayletterbadge{3}). 
At each step, it provides explicit evidence that rules were followed before it moves on (Figure \ref{fig:system} \grayletterbadge{4}\grayletterbadge{5}). 

Enforcement is driven by a combination of \texttt{ZORO.md} and \texttt{zoro-cli}. \texttt{ZORO.md} is generated on initialization and prepended to the repository's existing rules file (e.g., \texttt{AGENTS.md}), ensuring the agent loads it at the start of every session. Rather than a sprawling set of guidelines, \texttt{ZORO.md} gives the agent two CLI commands: \texttt{zoro~update-step} and \texttt{zoro~prove-rule}. We rely on the fact that coding agents always read their own terminal outputs; thus every CLI response is a reminder of \texttt{ZORO.md}'s protocol: a successful command tells the agent what to do next, and a failed one tells it exactly what it missed and what it must do to fix it (e.g. if the agent marks a step complete before enforcing all strict rules, the CLI rejects the command and returns an error listing what rules it needs to prove). Since the CLI output re-encodes the protocol at every step, it prevents \texttt{ZORO.md} from context rot.  
\begin{figure}[h]
    \centering
    \includegraphics[width=.5\textwidth]{figures/system_interface_rule_refine.png}
    \caption{Batch Rule Refine Modal. \change{Users click \textit{Evolve} to access this modal and iterate, edit, and approve all the refined rules.}}
    \label{fig:system_interface_rule_refine}
\end{figure}

\subsection{Evolving \change{the Scope of the Rules}}

Because the evidence is on the enforcement panel, Johnny can evaluate not just whether a rule was followed, but whether the rule itself was specific enough. While reviewing the evidence for \texttt{Rule B}, he realizes the migration ran correctly, but the rule was too vague to capture what he actually wanted. He clicks directly on the enforcement evidence and adds a note:

\vspace{0.5em}
\noindent
\fcolorbox{gray!60}{white}{%
\parbox{\dimexpr\linewidth-2\fboxsep-2\fboxrule\relax}{%
\scriptsize\fontfamily{phv}\selectfont
this worked but migrations should live in the migrations folder\dots also it didn't tell me how to run it. maybe also check that prior schema relationships didn't break?
}}
\vspace{0.5em}

His feedback is immediate and grounded in a specific artifact (Figure~\ref{fig:system_interface}\grayletterbadge{H}). As the session continues, he annotates other rules the same way — flagging gaps, adding expectations, capturing implementation details he wants to encode going forward. 
The cycle of Codex writing code, submitting proof, and Johnny reviewing evidence continues through Steps \textit{1.2} through \textit{3}. Some steps move quickly because their rules are non-strict; others are slower due to enforcement. By the end of \textit{Step 3}, Johnny has shipped the feature and has a clear record of how every rule was applied along the way.

After Codex completes the plan, Johnny sees a \textit{Rule Review} section. 
It functions like a code review, except instead of reviewing code, Johnny reviews where his rules were applied and where they fell short. He clicks \textit{Evolve Rules}. \appname\ takes all accumulated notes and proposes updated rule definitions that incorporate them (Figure~\ref{fig:system_interface_rule_refine}\grayletterbadge{I}), which \appname\ renders in a diff view. Johnny steps through each rule and can accept a proposal as-is, edit it inline before accepting (Figure~\ref{fig:system_interface_rule_refine}\grayletterbadge{J}), or reject it entirely (Figure~\ref{fig:system_interface_rule_refine}\grayletterbadge{K}). Only accepted changes are saved back to the ruleset (Figure~\ref{fig:system_interface_rule_refine}\grayletterbadge{L}). The rule refinement for \texttt{Rule B} is seen on Figure~\ref{fig:system_interface_rule_refine}. It details the exact folder future migrations should live in and also details to prompt the user about the migration that Johnny had written in his in-situ  notes. 
What started as a loosely maintained markdown file has become living and evolving  — rules get more precise and more trustworthy every time Johnny and Codex build something together.



\subsubsection{Implementation}
\appname's evolve feature works as follows: 
In-situ notes are saved directly to \texttt{logpose/.zoro}, attached to the specific enforcement evidence they annotate (see Figure \ref{fig:system} \grayletterbadge{2}). 
When a user clicks ``\textit{Evolve},'' \appname\ loads all in-situ notes from \texttt{logpose/.zoro}, groups them by rule, and sends them alongside the current ruleset to an LLM for rule refinement (see Figure \ref{fig:system} \grayletterbadge{3}\grayletterbadge{4}). Once users approve the changes, the new rules are saved to the ruleset in \texttt{logpose/.zoro} (see Figure \ref{fig:system} \grayletterbadge{5}\grayletterbadge{6}\grayletterbadge{7}). In \appname, users can view rules and manage them in the \textit{Rule Management Tab}, which can be found in Appendix \ref{app:system_rulemanagement}.

\section{Technical Evaluation}

We begin with a controlled evaluation to determine if \appname\ improves a coding agent’s ability to follow rules. Here, we isolated the impact \textit{Enrich-Enforce} in \appname\  before testing the full system (\textit{Enrich–Enforce–Evolve}) with users. We ask: does \appname\ meaningfully improve \textit{how} agents implement code without degrading \textit{what} is being coded? For example, if the coding task is to create a start button, the \textit{what} refers to if the button exists and functions as intended, while the \textit{how} refers to if the button was implemented following the task’s development rules—such as using the design system Button instead of an external component library. \appname\ is considered successful if it is 1) equally effective at producing the feature (\textit{what}) and 2) is better at following the rules (\textit{how}). 

\subsection{Method}
We evaluated a total of 36 vibe coding sessions across 9 implementation tasks and 4 conditions. Tasks were drawn from two full-stack repositories developed from scratch: \textit{NoteSense} and \textit{SnakeGame}. 
\textit{NoteSense} is a full-stack note-taking and sensemaking application (22k lines of code) built with a TypeScript/React frontend, Python Flask backend, and SQLite database -- integrating with OpenAI, Google Tasks, and Gmail for AI-assisted organization, task synchronization, and email export. It represents a standard full-stack project. \textit{SnakeGame} is a full-stack browser game (11k lines of code) built with a TypeScript/React frontend and Node.js/Express backend and a SQLite database. It demands more complex UI concerns: real-time game state management, continuous rendering, interactive graphics, and dynamic user input. It also incorporates accessibility testing via jest-axe and component development through Storybook. Both repositories and project-specific rules were built from scratch, giving us full visibility into design and architectural decisions.


We evaluated five tasks for \textit{NoteSense} and four for \textit{SnakeGame}. Each task represents a substantial, multi-step development problem spanning 500–5,000 lines of code, rather than on short, isolated tasks that models can solve easily. Tasks cover full-stack development, including UI components, database logic, system integration, and AI-assisted functionality (see Appendix \ref{app:techeval_tasks}) \change{that reflect real-world challenges identified from exploratory interviews in Section 3}. \change{Our evaluation included one run per task; our goal was breadth rather than collecting repeated data points for fewer tasks.}

We used Cline as the coding agent with Claude Sonnet 4.6 as the model for our technical evaluation.
Cline\footnote{https://github.com/cline/cline} is an open-source, model-agnostic tool widely adopted in professional development workflows, with ~60k GitHub stars. It was chosen for this evaluation for two reasons: 1) being open-source, its system prompt is fully transparent — unlike tools such as Claude Code or Codex, where the underlying system prompt is hidden; 2) it provides detailed execution logs showing exactly where the agent modifies code within the repository, enabling annotators to post-hoc verify whether development rules were followed. 
Claude Sonnet 4.6 was selected as a strong, general-purpose coding model.
\change{Our evaluation is scoped to Cline; however \appname's enforcement mechanism operates as a model- and agent-agnostic self-verification layer. Because prior work has shown that self-verification techniques generalize across LLMs~\cite{weng2023largelanguagemodelsbetter}, we expect the enforcement layer to likewise transfer across coding agents.}

\change{In our current implementation, the full \textit{Enrich–Enforce} workflow increases token overhead because Cline must use the CLI to self-verify and mark steps as complete. In our evaluation, this increased token usage by approximately 4× compared to standard vibe coding. We revisit this tradeoff in Section~9.}


\subsubsection{Conditions}
To isolate the contributions of \appname’s components, we defined four experimental conditions:
\begin{itemize}
    \item \textit{No \appname\ (Baseline)}: Standard vibe coding. Agents execute tasks based on a normal plan. All rules are stored in \texttt{AGENTS.md}, but they are not actively integrated into the plan.
    \item \textit{Basic \appname} \textit{(No Enrich, No Enforce)}: Agents mark steps as complete and rules exist in \texttt{AGENTS.md}, but they are not used to enrich the plan and the agent does not have to provide proof that they followed each rule. 
    \item \textit{Partial \appname} \textit{(Enrich, No Enforce)}: The plan is enriched with rules, and the agent must mark steps as complete, but does not have to provide proof that they followed each rule.
    \item \textit{Full \appname} \textit{(Enrich, Enforce)}: The plan is enriched and rules are enforced, and the agent must mark steps as complete. The agent cannot complete steps until all rules are followed.
\end{itemize}

For all conditions, we provided the task description and development plan, after which the agent generated code independently. If compiler errors occurred, we applied minimal patches to enable compilation and recorded the number of patches applied. The resulting implementations were then evaluated.

\subsubsection{Metrics}
Because tasks vary in features, systems, and constraints, a single generic rubric would fail to capture task-specific requirements. Thus, we define a specific rubric for each task. Each rubric specifies what must be implemented (feature completeness, the \textit{what}) and which task-specific development rules must be followed (rules followed, the \textit{how}):

\begin{itemize}
    \item \textit{Feature completeness} measures whether the intended functionality was implemented and usable within the system. Partial implementations (e.g., backend logic without a usable front-end interface) were scored as incomplete, since the feature could not be tested within the application. 
    \item \textit{Rules followed} measures whether the agent adhered to task-specific rules. \change{Each rule contributes a single pass/fail outcome per session.} 
    If a rule was applied multiple times and was violated in any instance, it was marked as not followed. This scoring approach prioritizes consistent adherence and provides a conservative estimate of rule following. 
\end{itemize}

Two annotators, both software engineers with at least five years of computer science experience, independently graded each task using these rubrics. Annotators reviewed both the generated code and the associated vibe coding chat history, since some rules govern interactive behaviors (e.g. checking in with the user before installing a new package) rather than code alone. Annotators were compensated \$200 each (\$40 an hour).  

\subsection{Results}
We first verify annotation consistency between the two raters. Using \change{Cohen's Kappa}, we observe strong agreement across both \change{\textit{rules followed} (corr=0.97) and \textit{feature completeness (corr=0.82)}}, suggesting that our annotations are reliable for downstream analysis. 

\begin{table}[h]
  \centering
  \small
  \setlength{\tabcolsep}{4pt}
  \begin{tabular}{lll}
    \toprule
    Condition & \makecell{Feature Completeness \\ ($\mu$, 95\% conf)} & \makecell{Rules Followed \\ ($\mu$, 95\% conf)} \\
    \midrule
    No \appname & 0.64±0.144 & 0.51±0.08 \\
    Basic \appname & 0.65±0.17 & 0.55±0.11 \\
    Partial \appname\ & 0.61±0.11 & 0.68±0.07* \\
    Full \appname\ & 0.77±0.13 & \textbf{0.80±0.08*} \\
    \bottomrule
  \end{tabular}
  \caption{\textit{Feature Completeness} and \textit{Rules Followed} Avg. Scores.}
  \label{tab:techevalresults}
  \vspace{-12pt}
  \centering \small * denotes significance \change{($p < 0.05$)} compared to No \appname\ \textit{(Baseline)}
\end{table}

\subsubsection{Rules Followed was highest in Full \appname}
Overall, full \appname\ had more rules followed than the other conditions, with a score of 0.80±0.08 (µ, 95\% conf). This result aligns with our expectations, as the Full condition combines all system capabilities.
A one-way ANOVA reveals a statistically significant difference across conditions for \textit{rule following} ($p<0.05$). Post hoc pairwise paired t-tests with Holm-Bonferroni correction showed that Full \appname\ significantly outperformed all other conditions ($p<0.05$), and Partial \appname\ significantly outperformed No \appname\ ($p<0.05$). All other pairwise differences are not statistically significant. \textbf{This pattern suggests that individual components show some improvements in the agent’s ability to follow rules, but \appname’s largest gains are when enrichment and enforcement are combined.}


Notably, even in the best condition, rules are followed only 80\% of the time. A qualitative analysis of the rules suggests that failures often stem not from the agent’s inability to follow the rules, but from limitations in the rules themselves. Rules frequently contain ambiguities, gaps, or unintended loopholes. Agents tend to follow rules “to the letter,” which can lead to behavior that satisfies the literal text while diverging from the intended human meaning (see Appendix \ref{app:techeval_vaguerules}), suggesting that static rulesets are insufficient for reliable \change{rule following}. To achieve true consistency, rules must be treated as evolving artifacts – continuously refined as misinterpretations and edge cases emerge. 

\subsubsection{Feature Completeness in \appname\ is on par with standard vibe coding}
We observe similar levels of \textit{feature completeness} across all conditions. Full \appname\ achieved the highest score 0.77±0.13, which is 0.12 higher than any other condition. However, a one-way ANOVA reveals that these differences are not statistically significant. While a larger sample size may reveal a bigger trend, it is more important that rules are followed without dimming feature-completeness. Our results indicate that \appname\ ’s value remains centered on how a task is performed (rules) rather than what is implemented (features). 



\subsubsection{What types of rules does \appname\ work best for? }
We investigate which rules benefit most from enforcement. Our data reveals a meaningful distinction between two categories: \textit{coding rules} and \textit{process rules}. \change{These categories were derived through an iterative thematic analysis of the rules, with one author performing open coding and the research team refining categories until consensus.} \textit{Coding rules} govern how code should be written — for example, naming conventions, API structure, or formatting patterns that are visible directly in the codebase. \textit{Process rules}, by contrast, govern what the agent must do while writing code, such as running a migration after a schema change or requesting confirmation before a destructive action. These rules encode explicit human intent that cannot be inferred by looking at the code alone. 
For process rules, rule following rises from 35\% in standard vibe coding to 87\% in the full system. For coding rules, the gain is more modest, from 59\% to 75\% (see full table in Appendix \ref{app:techeval_codingandprocessrules}). This makes intuitive sense: coding agents can already pick up on coding rules by recognizing patterns in the existing codebase, and as models improve, enforcement may become less necessary. \textbf{Process rules, however, capture intent that leaves no trace in the code itself, making both enrichment and enforcement essential.}

\section{User Study}

We conduct a user study with the full \appname\ system (\textit{Enrich-Enforce-Evolve}) to explore the following research questions (RQs): 

\begin{itemize}
    \item \textbf{RQ1}: What benefits does \textit{Enrich-Enforce-Evolve} provide?
    \item \textbf{RQ2}: How does \appname\ change users' vibe coding behaviors?
    \item \textbf{RQ3}: How does \appname\ change users' cognitive strategies when vibe coding?
\end{itemize}

\subsection{Participants}
We recruited twelve participants (7F, 5M) via word-of-mouth and snowball sampling. All participants \change{were expert vibe coders} who had at least five years of coding experience; \change{thus, participants could meaningfully compare \appname\ against their existing vibe coding practices.} 10 participants used rules files on a daily basis. Four participants were PhD students across technical fields, and the remaining eight were industry software engineers or data analysts. 
Participants were compensated \$50 for their time.
\subsection{Procedure}
Following approval from our Institutional Review Board (IRB), we conducted \change{a think-aloud study}. Each participant completed a 90–120 minute session, where they used rules to vibe code two tasks using \appname. For consistency, Codex was chosen as the coding agent, which all participants were familiar with. \change{Since \appname\ is agent-agnostic, similar to the technical evaluation, we expect our findings to generalize across agents, although our results only scope to Codex}.
Each session followed a consistent structure:
\begin{itemize}
    \item \textit{Onboarding (15 min)}. Sessions began with a walkthrough of the core features on \appname's interface.
    \item \textit{Setup and Repo Selection (15 min)}. The facilitator helped them set up \appname\ and upload their ruleset from an existing \texttt{AGENTS.md}. Participants chose to vibe code between two setups: a repository and rules they brought themselves, or \textsc{Poneglyph}, a full-stack journaling app with a ruleset that we provide, which they were asked to add rules to.
    \item \textit{Brainstorm and Complete 2 Tasks (30–45 min per task)}. Participants brainstormed their first task; after completing the first task, they brainstormed a second task in the same repository, optionally added more rules they wanted in their ruleset. Tasks were self-contained and representative of everyday development work — not trivial, but scoped to what a developer could realistically tackle in a session. For both tasks, participants were instructed to think aloud. After \appname\ generated an initial plan, participants reviewed the rules and selected which to strictly enforce (and test). They then completed the task with Codex and \appname, reviewed enforcement evidence, and wrote in-situ notes. After each task, participants performed batch rule refinement.
    \item \textit{Interview (15 min)}. Each session concluded with a semi-structured interview exploring participants' experiences with \appname, their mindset and behaviors around vibe coding with active rules, and reflections on how the tool compared to their existing workflows. \change{These reflections are self-reported and may be subject to recall bias. We therefore interpret our findings as expert's perceptions of workflow changes rather than objective measurements of behavioral change.}

\end{itemize}


Seven participants brought their own repositories (research or personal projects), while five used \textsc{Poneglyph} (see Appendix \ref{app:userstudy_participants}). Notably, participants who brought their own repositories treated the study as a real working session, intending to actually use the code produced. 
\change{An author conducted open coding of notes and transcripts, iterating on emerging codes during research meetings with the broader team to identify recurring themes.}

\section{User Study Results}
In total, participants enforced 166 rules (110 unique). 10-15 rules were included in each plan, and participants enforced an average of 9.17 rules per session. Participants who trusted agents tended to enforce fewer rules, while skeptical participants enforced more. 10/12 participants reported they found it helpful that the rules and enforcement evidence were visible on the interface; the other two found it useful but felt a more compact representation of the same information would be better. \change{Five participants reported increased cognitive load due to the text-heavy experience and having to switch between the coding agent and \appname's interface.} \change{Nine wanted rule enforcement directly embedded in VSCode, such as inline highlighting of rule evidence within the coding agent itself.}  10/12 participants made use of the in-situ notes to \textit{evolve} rules; the remaining two did not because their rules were well-specified enough to not require refinement. 
All 12 participants said they would use a system like \appname\ in their everyday workflows.

Rule enforcement improved from Task 1 to Task 2. 
Task 1 had a pass rate of 77.9\% (60/77) and task 2 had a pass rate of \change{94.4\% (84/89). Participants reused 56 rules between the tasks. 43/56 rules were applied correctly in task 1; 13/56 were refined after task 1 because they were incorrectly applied. After refining these rules, 12/13 of these rules were correctly applied in task 2; only 1/13 was refined again after task 2, supporting our hypothesis that underspecified rules benefit from the \textit{Evolve} mechanism.}

\subsection{RQ1: What benefits does \textit{Enrich-Enforce-Evolve} provide?}
\subsubsection{Enrich + Enforce makes vibe coding auditable}
\change{Nine participants reported that \appname\ improved their ability to inspect agent behavior, and seven had debugging instances where enforcement evidence helped them understand agent behavior.}
They contrasted \appname\ with standalone vibe coding, where the experience amounted to, as P5 described, "staring at an empty window and hoping for the best" while the agent produced dozens of changed files with little explanation. With \appname`s enriched plan and enforcement evidence, "you are more aware of what is happening in the background — you know what is being applied versus not applied" (P4). P12, an ML engineer translating her research paper to code, required the agent to cite how each line of code mapped to variables for each math equation in her paper. The enforcement evidence allowed her to trust the code generated was actually following her math, and she explained: "when something fails, it is helpful to know that the implementation is faithful, so you know the issue is with [my] algorithm rather than with a [coding] bug." 

This auditability also allowed participants to debug their rules. P6, who generated heatmaps from experimental data, hit a bug where his outputs were not loading. He traced through the enforcement evidence and found a rule that contained an incorrect path to a font file. Without \appname, he would have otherwise had to, as he described, "prompt over and over again... you can't even remember the rules you want the agent to follow." P3 concluded, "either way, I'm going to look for a way to verify the usage of these rules, so having these explicitly visualized saves me a lot of time."  


\subsubsection{Enforcement as selective controls.}
\change{All participants enforced at least three rules per task; on average each participant enforced 9.17 rules}. Participants primarily chose to enforce: 1) rules whose violation caused compounding downstream failures (type safety, compiler checks, testing), 2) compliance and security rules where hallucinated outputs carried real-world consequences, 3) recurring personal preferences they re-specified prompt by prompt. 

Participants did not enforce the same rules across both tasks. As they observed rules being followed, they began extending more trust to the agent. Of the 77 rules enforced in Task 1, 56 (72.7\%) carried over to Task 2.  P1 chose not to re-enforce rules like "import icons only from the icon registry", trusting that the rule being explicitly mentioned in the plan was sufficient, stating: "I didn't enforce the [rules in task 2] that were demonstrated previously [in task 1] were correct."  
P3 described a threshold where, after seeing a rule pass, she would stop enforcing it, "as long as [she had] a way to see a historical base of when it has been used."

\subsubsection{Evolve makes rules dynamic}
\appname\ allowed participants to precisely edit their rules; \change{10/12 participants evolved at least one rule.}
To P1, \textit{Evolve} made rules "a dynamic thing...[allowing you to] go more specific until you get what you want". Users only realize a rule is underspecified once they've seen it fail. P3 had a rule prohibiting exposing personally identifiable information (PII) to users. However, the agent omitted the \texttt{customer\_uuid} field entirely thinking it was PII. P3 resolved this by adding an in-situ note with an allowlist defining what did and did not constitute PII in her context. Vague rules require iterative tightening -- as P11 put it, "if you are not really clear, [the LLM] just finds a loophole to "do" what you asked without actually doing it." 
Rather than regenerating the rules or appending them to the rules file, an approach P8 described as getting "messy very quickly" because "you lose track of what changed and why", \appname's in-situ notes allowed participants to "modify rules in-place," keeping changes targeted and traceable. 

\subsection{RQ2: How does \appname\ change users' vibe coding behaviors?}
Based on their prior experience with coding agents, nine participants were initially skeptical that coding agents would follow their rules. After using \appname\ to enforce their rules, however, they became more confident. The remaining three participants already trusted agents, but still appreciated how \appname\ turned rules into active context they could visualize. As P6 put it: "if [\appname] is validating and showing that it is following my rules then I am way more happy to put my trust in the system"
P11 similarly described trust being built through "proof that it happened." 


A direct consequence of this confidence was a marked reduction in supervision overhead. 
P5, an ML PhD student, described typically breaking his workflow into discrete verification stages — checking tensor data types, device placement, component modularization, and using dummy data to test procedures at each step— because he didn't trust an agent to maintain these constraints end-to-end. In our study, he encoded them directly as rules 
and let \appname\ enforce them in his place. The result was that his usual multi-stage verification process was externalized into the system itself. Seeing every rule surface and pass broke down his habitual skepticism: "All the rules passed and I'm shocked because I'm usually the manager and I don't trust it." P11, vibe coding a rich text editor on \textsc{Poneglyph}, encoded a rule to not write unnecessary try-catch statements — a constraint he had little faith the agent would respect. Seeing the enforcement evidence surface and confirm the rule had been followed allowed him to visibly relax, stating "I would probably need something like this in bigger projects because it will get more chaotic — and if you have the proof that it followed the rules, then you don't have to look inside the codebase and check each output."


\subsection{RQ3: How does \appname\ change users' cognitive strategies when vibe coding?}
Nine participants developed a strategic shift when vibe coding with \appname\ — away from "what should I say in this prompt?" and toward "what rules should govern my plan?" A key reason is that rules and prompts operate at different scopes: a prompt must be re-specified every session, while a rule, once written, persists and compounds across future sessions. P8 articulated this the most explicitly:
\begin{quote}
\textit{"It kind of is a different framework of developing because you end up being not only plan-focused, but now rule-focused as well... if I live in a world where I can see what rules are applying to my plans, and I can figure out that this rule is leading this plan astray, or this rule is really important to this plan — I can really think of it from a mindset of: these are the rules I need for my plans to operate 10, 20, 30 times more efficiently."}
\end{quote}
Rather than treating a ruleset as a one-time setup cost, participants began reasoning about rules as the primary mechanism to compound the agent's long-term usefulness. P7 connected this to a broader problem about the quality of vibe coding in general: 
\begin{quote}
    \textit{"One concern with vibe coding is that you get loosey-goosey, but that loses the structure you need in traditional coding. Introducing rules at the forefront brings traditional coding back. You are thinking more strictly as a coder rather than just having a casual conversation."}
\end{quote}

This shift toward rules-first thinking raises a natural question: does enforcing rules constrain the agent's creative autonomy? In our study, participants developed a meta-strategy of enforcing rules where correctness mattered, and relaxing enforcement where they wanted the agent to exercise judgment. P8 referred to it as rules to be followed "to the letter" versus those to be "inspired by." For example, P6 chose not to re-enforce his \textit{"Use MATLAB Parula colormap by default"} rule in Task 2, reasoning that for a new type of heatmap, the agent might independently select a better option. P7, who frequently plots data, left rules around SQL queries and backend packages unenforced "because I wanted the agent to have flexibility on what it thought was best." He instead reserved strict enforcement for rules about plot appearance, where he had strong preferences. This selective enforcement suggests that rules did not feel uniformly constraining but instead provided a way to deliberately choose between control and agent autonomy.

Overall, participants were struck by the breadth of rules \appname\ could successfully enforce — and for many, seeing it work in practice shifted their baseline assumptions about what AI agents could be trusted to do autonomously. 

\section{Discussion}
\subsection{From Prompt Engineering to Rule Engineering}

In vibe coding, the path of least resistance is to click approve agent suggestions, defer verification, and accumulate technical debt that surfaces later. Agents code at the speed of light, and it is only human to keep pace by accepting outputs without scrutiny. But for developers who care about \change{their rules being followed}, this tradeoff is costly; code bases can quickly become messy and unrecognizable. \change{Our user study shows that \appname\ shifts cognitive load from repeatedly checking code outputs to auditing rules. Although this requires effort, it is an investment that accumulates because developers don't need to micromanage the coding agent in the future.}  

In our evaluations, we saw a broader shift in the role of humans from prompt engineers to active shapers of agent behavior.
Rules are not merely prescriptive instructions about low-level details (e.g. what color a button should be, or how a function should be named), but can be process rules that encode decisions around agent autonomy, permissions, and oversight that will always require human judgment.
Rules are ultimately a type of policy: something that can be applied broadly to govern behavior across context. 
\textit{Enrich-Enforce-Evolve} operationalizes this in vibe coding, but the same principle applies across any agentic workflow. 
A rule that requires checking for an existing Jira ticket before creating a new one, or that mandates a summary be sent after every meeting, encodes organizational process just as naturally as a rule that forbids installing unvetted dependencies. 
The paradigm shift, then, is this: if the last decade of software development was defined by the question of how to write better code, the next may be defined by the question of how to write better \textit{rules}. Steering agent behavior through explicit, enforceable, and evolvable policies, rather than relying on implicit model behavior or ad-hoc prompting, may prove to be the defining human skill of the agentic era.

\subsection{Passive Context to Active Enforcement}
When designing \appname, we observed that developers did not mainly struggle to express their intent; they struggled to trust their intent would be followed. Thus, the future of vibe coding will likely demand not just better rules, but better guardrails.  Test cases are one such guardrail; they verify what an agent produces. Rule enforcement is another; it lets users control what an agent can do. Together, they form a complementary scaffolding that grants agents greater autonomy because their behavior is bounded and verifiable.



In our current design, enforcement is grounded in unit tests and evidence traces. Participants in our user study suggested it could extend further: dedicated enforcement agents, multimodal checks over UI and system behavior, or validation of agent reasoning traces to ensure the agent considered specific constraints before acting. More broadly, enforcement need not rely on LLMs at all — typed interfaces, deterministic program logic, regex-style validators, and domain-specific policy engines are all viable mechanisms depending on the context. In this sense, \appname\ is not just a system that enforces rules, but a foundation for treating policy enforcement itself as a design primitive in agentic systems.


\section{Limitations}
Our user study involved 12 sessions across two tasks  with expert vibe coders. The tasks operated on single-developer-style codebases, and participants completed two tasks each; longitudinal studies with more diverse participants and a broader range of tasks would provide a richer understanding of real-world adoption. 

Our technical evaluation was limited to 36 total cases due to the complexity of our tasks and the high computational cost of running the full enforcement pipeline end-to-end. Enforcement introduces non-trivial overhead, increasing token usage by approximately 4× due to additional agent actions required. \change{Future iterations of \appname\ could reduce this overhead through prompt caching and by incorporating deterministic enforcement mechanisms, such as static analysis, Git hooks, and linters for rules that do not require LLM reasoning. These improvements would lower token costs and enable larger-scale evaluations.}


\change{A broader limitation is that enforcement evidence is not a formal guarantee a rule was followed — developers who over-trust it risk false confidence. Its real value, as our user study showed, is making rule-following transparent and auditable. Pairing this with the deterministic enforcement mechanisms discussed could offer stronger guarantees for rules where evidence alone is insufficient.}

At the system level, \appname\ currently performs static rule retrieval: \appname\ loads task-relevant logic once at initialization and cannot autonomously fetch updated rules or adapt its strategy in response to real-time state changes. Future work will explore dynamic retrieval mechanisms to enable more responsive, context-aware behavior. 
\appname\ is implemented as a Python package requiring manual setup. Participants felt that a lighter-weight form factor, such as a VSCode extension \change{or in-line highlighting of enforcement within the coding agent window} could significantly improve \change{cognitive load}. 
\change{Some participants were also skeptical of AI-driven enforcement. They noted that code snippets and unit tests are insufficient evidence for certain rule types (like performance or UI rules) and may instead need enforcement in live browser environments. 
}


\change{Finally, overall ruleset management remains an open challenge.} While we introduced merge functionality to consolidate rules, \change{\appname\ does not yet address questions such as resolving conflicting or contradictory rules, organizing large rule sets, and maintaining them over time.}
We can also introduce better methods for rule creation; while \appname\ handles this by a structuring rules from a pre-existing \texttt{AGENTS.md} file (see Appendix \ref{app:system_rulemanagement}, \ref{app:system_setupandruleimport}, \ref{app:system_monitoring}), more work can be done to support users. 

\section{Conclusion}
Vibe coding has shifted the developer's role from writing code to steering agents — but without mechanisms to enforce intent, that steering remains untrustworthy. \appname\ addresses this by transforming rules files from passive documentation into active controls through the \textit{Enrich-Enforce-Evolve} pattern.
Our technical evaluation shows that agents follow rules more reliably under this paradigm, and our user study reveals how active rules change the experience of vibe coding itself.
More broadly, \appname\ reframes what it means for a developer to interact with rules. Rather than dropping constraints into a text file and hoping they stick, rules become controls developers can see, interact with, and trust.
As agents grow more autonomous, the question of how humans maintain meaningful control becomes increasingly important. \appname\ shows that it will require more than prompting; it will require mechanisms that make developer intent visible, enforceable, and adaptable over time.

\begin{acks}
\change{
We thank Eugene Wu, Omar Shaikh, Vivian Liu, and  Dora Zhao for their thoughtful feedback on earlier drafts of this paper. We are also grateful to Faris Ibrahim, Parmida Shariat, Sebastian Salazar, Brenda Borbon, Amy Huang, Kenny Chen, Nikos Pagonas, and Angela Ma who provided ongoing support through discussions. Finally, we thank our study participants for their time and insights.
}
\end{acks}

\bibliographystyle{ACM-Reference-Format}
\bibliography{sample-base}

\onecolumn

\appendix
\section{System}

\subsection{Library and Frameworks}
\label{app:system_libraryandframeworks}
\appname’s interface is written in Typescript and React. The backend is written in Python and Flask. 
The backend handles reads and writes to the \texttt{.zoro directory} to coordinate communication between the interface and the agent. 
For plan enrichment, rule refinement, and supervision, \appname\ uses GPT-5.3, leveraging its extended context window to reason across long-range dependencies.

\subsection{Setup and Rule Import}
\label{app:system_setupandruleimport}
Johnny initializes \appname\ in his repository by running \texttt{zoro init -{}-user-name Johnny} in the root directory. This creates a \texttt{.zoro} folder which serves as the shared workspace between \appname\ and the coding agent. He then starts \appname\  by running \texttt{zoro -{}-api}, which launches the \appname\  frontend in a local browser window.

To import his existing rules, Johnny navigates to the Rule Management tab and clicks the button to structure his existing \texttt{AGENTS.md} file into rules. \appname\  parses the file, transforming each information block into a structured JSON ruleset saved to \texttt{.zoro}. The rules are immediately displayed in the interface — in Johnny's case, approximately 50 rules are extracted and ready to be reviewed, categorized, and assigned enforcement levels before his first session.

\subsection{Example Plan and Evidence}
\subsubsection{Example Initial Plan Generated by Codex}
\label{app:system_initialplan}
Below shows the example plan generated by Codex.

\noindent
\begingroup
\setlength{\fboxsep}{1mm}
\setlength{\fboxrule}{0.5pt}
\fcolorbox{gray!50}{white}{%
\begin{minipage}{\dimexpr\linewidth-2\fboxsep-2\fboxrule\relax}
\scriptsize\sffamily

\textbf{Original Agent Plan}

\textbf{Step 1: Category System Foundation}\\
Step 1.1: Add \texttt{Category} model to the backend schema (default category color: grey); Add \texttt{category\_id} field to \texttt{LogEntry} records\\
Step 1.2: Create a \texttt{Category} service in \texttt{api/categories.py} to support CRUD operations\\
Step 1.3: Implement a category sidebar view and integrate it with the main log entry interface\\[0.5em]

\textbf{Step 2: AI Auto-Organization...}\\
Add a method to cluster log entries by sentiment using fuzzy string matching\\
Create a \texttt{/preview} API endpoint to return AI categorization suggestions\\
Implement logic to apply the selected organization\\[0.5em]

\textbf{Step 3: Preview Dialog and Frontend...}\\
Display category suggestions in expandable sections\\
Support drag-and-drop adjustments for managers to refine groupings

\end{minipage}%
}
\endgroup

\subsubsection{Example Enforcement Evidence for \texttt{Rule B}}
\label{app:system_enforcementevidence}
Below shows the example enforcement evidence for \texttt{Rule B}.

\begin{GrayPanel}
\textbf{Step 1.1: Add Category Model to backend schema}\par
\scriptsize
$\bullet$ Ask the user to choose the default category color\par
$\bullet$ Add \texttt{category\_id} to \texttt{LogEntry} records\par
$\bullet$ Backfill existing data safely\par

\RuleBox
  {RULE A [strict]: Always ask the user before making design decisions.}
  {User said ``Default green color that is used in the header in the frontend''}

\RuleBox
  {RULE B [strict] [testable]: Ensure all schema changes properly backfill or migrate existing data.}
  {Created migration script in \texttt{backend/migrations/...}}

\textbf{\textsf{\scriptsize Code Evidence}}\par
\begin{lstlisting}
# Run: flask db upgrade
def upgrade():
    for entry in LogEntry.query.filter_by(category_id=None).all():
        entry.category_id = DEFAULT_CATEGORY_ID
    db.session.commit()
\end{lstlisting}

\textbf{\textsf{\scriptsize Test Case}}\par
\begin{lstlisting}
def test_category_backfill():
    assert all(e.category_id is not None for e in LogEntry.query.all())
    assert Category.query.get(DEFAULT_CATEGORY_ID).color == 'color-primary-500'
\end{lstlisting}

\noindent
\fcolorbox{green!60}{green!10}{%
  \parbox{\dimexpr\linewidth-2\fboxsep-2\fboxrule\relax}{%
    \scriptsize\fontfamily{phv}\selectfont
    \textbf{Status:} Passed\par
    \textbf{Execution Time:} 0.42s\par
    \textbf{Runs:} 1
  }%
}
\end{GrayPanel}

\subsection{\texttt{ZORO.md} file}

\begin{lstlisting}[
    language=bash, 
    breaklines=true, 
    basicstyle=\scriptsize\ttfamily,
    frame=single,
    caption={ZORO Protocol Markdown Content}
]
# ZORO Protocol
This file defines the required execution protocol for agents.

## Core Rule

Before starting work, read `.zoro/zoro_plan.md`.

For every plan step, the agent must:
1. Call `zoro update-step <step-id> in_progress` when work begins.
2. Call `zoro prove-rule <step-id> --rule "..." --evidence "..."` for each required rule.
3. Only after required rule evidence is submitted, call `zoro update-step <step-id> completed`.

## Hard Stops
- Do not mark a step `completed` before required `zoro prove-rule` calls.
- Do not start the next step while the current step is incomplete.
- If evidence is missing or weak, remain on the current step and add better proof.

## Evidence Guidance (Concise)
Evidence should be specific and audit-friendly:
- What changed
- Where it changed (file/component)
- Why it satisfies the rule
- Optional verification signal (test result, output, screenshot note)

## `prove-rule` Command Shapes

Use the non-test form when test evidence is not required:

```bash
zoro prove-rule <step-id> --rule "<rule text>" --evidence "<what changed and why it satisfies the rule>"
```

Use the test-evidence form for testable rules:

```bash
zoro prove-rule <step-id> --rule "<rule text>" --evidence "<what changed and why it satisfies the rule>" --test-name "<test_name>" --test-command "<command>" --test-result pass --test-output "<summary>" --test-file "<path>"
```

## Example

```bash
zoro update-step step-1-1 in_progress
zoro prove-rule step-1-1 --rule "Use repository pattern" --evidence "Added UserRepository in backend/repositories/user_repository.py and refactored UserService to depend on it."
zoro update-step step-1-1 completed
```

### Example With Test Evidence

```bash
zoro update-step step-2-3 in_progress
zoro prove-rule step-2-3 --rule "Strict testable rule: API parsing must handle nested response fields" --evidence "Updated parser in frontend/src/services/api.ts to read supervision.status and token_stats safely." --test-name "test_nested_response_parsing" --test-command "pytest tests/backend/test_api_parsing.py -k test_nested_response_parsing" --test-result pass --test-output "1 passed" --test-file "tests/backend/test_api_parsing.py"
zoro update-step step-2-3 completed
```
\end{lstlisting}

\label{app:system_zoromd}

\subsection{Supervision and Rule Learning}
\label{app:system_monitoring}
As Codex executes the enriched plan, Johnny monitors its progress through two collapsible floating windows accessible from the main interface: a \textit{Supervisor Panel} and \textit{Rule Learning Panel}. The \textit{Supervisor Panel} displays the current step, the rules being enforced, and a high-level status summary that updates in real time. For example, after Step 1.1 completes, the panel reads:
\vspace{0.5em}
\noindent
\setlength{\fboxsep}{1mm}
\setlength{\fboxrule}{0.5pt}
\fcolorbox{gray!60}{white}{%
\parbox{\dimexpr\linewidth-2\fboxsep-2\fboxrule\relax}{%
\ttfamily\scriptsize
On track --- Codex has completed step 1.1, verified 2 strict rules, and is moving on to step 2.
}}
\vspace{0.5em}

If the agent deviates from the protocol — skipping a step, failing to verify a rule, or becoming stuck — the panel surfaces the issue immediately. Johnny can pause the agent, redirect it, or prompt it to re-read \texttt{ZORO.md}. This provides a safety net for cases where the agent still deviates despite \texttt{zoro-cli} constraints.

The \textit{Rule Learning Panel} operates in parallel, surfacing candidate rules inferred from the current session. After Johnny confirms that new folder icons should default to green, \appname\ proposes a new rule that all new icons will have default color green. He accepts it, ensuring that future sessions will default to this behavior without requiring him to prompt for it again. Together, the two panels allow Johnny to monitor execution and incrementally refine his ruleset without leaving the interface.

\begin{figure}[h!] 
    \centering
    \includegraphics[width=\textwidth]{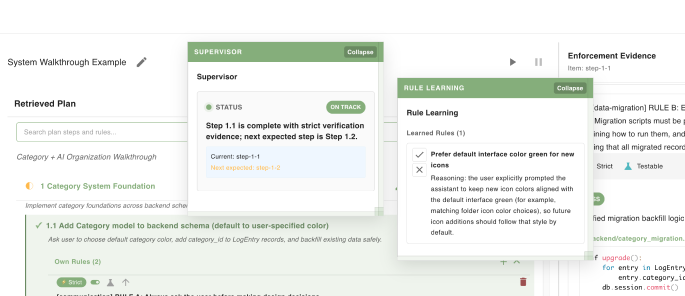}
    \caption{Supervisor and Rule Learning Panels}
    \label{fig:system_monitoring}
\end{figure}

The floating windows can be seen in Figure \ref{fig:system_monitoring}.

\subsection{Rule Review}
The Rule Review interface in the Visualization tab can be seen in Figure \ref{fig:system_rulereview}.
\label{app:system_rulereview}
\begin{figure}[h!] 
    \centering
    \includegraphics[width=\textwidth]{figures/system_interface_rule_review.png}
    \caption{Rule Review}
    \label{fig:system_rulereview}
\end{figure}

\subsection{Rule Management}
\label{app:system_rulemanagement}
The \textit{Rule Management} tab provides a structured interface for inspecting and organizing the full ruleset. Each rule is displayed as a card showing its title, description, category (e.g., \textit{ui}, \textit{backend}, \textit{workflow}), enforcement level, confidence score, and decay score. Confidence reflects how strongly a rule aligns with verified prior user behavior; decay captures how general or context-specific a rule is — higher decay indicates a narrowly scoped rule, while lower decay indicates one that applies broadly. When rules are first initialized from \texttt{AGENTS.md}, both scores are set to baseline values and update as rules are applied, refined, or merged over time. Inline controls allow Johnny to edit any rule, adjust its enforcement level, or remove it entirely.

On the right side of the interface, a set of organizational tools allows Johnny to consolidate his ruleset. He merges overlapping categories — for example, combining \texttt{ui} and \texttt{ui-ux} into a single unified category — and deduplicates similar rules, such as collapsing ``Buttons should be imported from the design system'' and ``Avoid importing buttons from external libraries'' into one precise, non-redundant rule. New rules can also be added manually at any time.

Learned rules from the \textit{Rule Learning Panel} and refined rules from the Evolve stage both appear here, keeping the full ruleset visible and editable in one place. This tab is not part of the primary session workflow; it is a persistent management surface that Johnny can return to at any time to audit, reorganize, and improve his rules independent of any active session. The \textit{Rule Management} interface can be seen in Figure \ref{fig:system_rulemanagement}.

\begin{figure}[h!] 
    \centering
    \includegraphics[width=\textwidth]{figures/system_interface_rule_management.png}
    \caption{Rule Management Tab}
    \label{fig:system_rulemanagement}
\end{figure}

\section{Technical Evaluation}
\subsection{Evaluation Tasks}
\label{app:techeval_tasks}

\begin{table}[H]
\centering
\small
\setlength{\tabcolsep}{3pt}
\begin{tabular}{|p{3.8cm}|p{9.2cm}|p{4cm}|}
\hline
\multicolumn{1}{|c|}{\textbf{Name}} &
\multicolumn{1}{c|}{\textbf{Description}} &
\multicolumn{1}{c|}{\textbf{Primary Technical Focus}} \\
\hline
AI-powered Note Organization & Automatically organizes notes into folders using LLMs, with chunking, folder preview, drag-and-drop, and grounded AI chat for querying notes. & Semantic content transformation, bulk state updates, human review before commit \\
\hline
Calendar Tracking & Displays note activity per day, week, and month, with day-view drill-down linking back to original notes. & Derived views over persistent state, aggregation, navigation across representations \\
\hline
Google Tasks Synchronization & Adds bidirectional sync with Google Tasks using OAuth and database migration, including previews, conflict resolution, and redundant-update avoidance. & Cross-system synchronization, authenticated API integration, conflict handling \\
\hline
Search and Version History & Supports search across current and historical notes, diff visualization, partial revert, and migration of existing notes into the new representation. & Temporal data modeling, schema evolution, reversible state changes \\
\hline
Email Export with LLM Assistance & Sends highlighted note sections to an LLM for feedback or rewriting, then integrates with Gmail for export using OAuth. & LLM-mediated transformation before external side effects, permissioned integration \\
\hline
\end{tabular}
\caption{NoteSense tasks and the distinct technical capabilities they exercise.}
\label{tab:notesense_tasks}
\end{table}

\begin{table}[H]
\centering
\small
\setlength{\tabcolsep}{3pt}
\begin{tabular}{|p{3.8cm}|p{9.2cm}|p{4cm}|}
\hline
\multicolumn{1}{|c|}{\textbf{Name}} &
\multicolumn{1}{c|}{\textbf{Description}} &
\multicolumn{1}{c|}{\textbf{Primary Technical Focus}} \\
\hline
Settings and Theme Management & Adds a settings page for fonts, colors, and themes via a color picker, along with named theme saving and a leaderboard/stats page. & Persistent user preferences, parameterized UI customization \\
\hline
Multiple Profiles & Supports creating and selecting profiles, each with its own settings, theme, and score history, while interacting correctly with the leaderboard. & Scoped persistence, profile isolation, per-entity state management \\
\hline
Gameplay Depth & Introduces configurable game variations such as power-ups, speed, and custom snake, head, and food designs, with settings grouped into levels. & Configuration-driven game logic changes, coupling UI state to runtime behavior \\
\hline
Game Hub and Profile Switching & Enables users to share and favorite themes and game designs in a hub, while applied copies remain local to the user. & Shared artifact browsing, copy-vs-local ownership semantics \\
\hline
\end{tabular}
\caption{SnakeGame tasks and the distinct technical capabilities they exercise.}
\label{tab:snakegame_column}
\end{table}
\subsection{Example of Agents Exploiting Vague Rules}
\label{app:techeval_vaguerules}
\begin{table}[H]
\centering
\small
\setlength{\tabcolsep}{3pt}
\begin{tabular}{|p{3cm}|p{3cm}|p{4cm}|p{6cm}|}
\hline
\multicolumn{1}{|c|}{\textbf{Rule}} & 
\multicolumn{1}{c|}{\textbf{Task}} & 
\multicolumn{1}{c|}{\textbf{What the agent did}} & 
\multicolumn{1}{c|}{\textbf{Why it failed}} \\
\hline
After a user action that changes state or data, the UI must reflect the change immediately without requiring a hard refresh
& Show updated sync status after Google Tasks sync completes in NoteSense
& Agent called \texttt{window.location.reload()} after sync completed. Sync status updated correctly after the reload.
& The rule states the UI must reflect changes ``without requiring a hard refresh'' — the user did not have to refresh manually, so the agent considered the rule satisfied. The intended behavior was to update local state or refetch in place, preserving scroll position and UI context. The rule's ambiguity around programmatic reloads left a loophole the agent exploited in a technically compliant but undesirable way. \\
\hline
Shared interactive components include Storybook stories (default and key variants)
& Add a theme preview component to the Game Hub in SnakeGame
& Agent created a single default story (\texttt{export const Default = () => <ThemeCard />;}) with no stories for locked themes, favorited state, applied state, or empty hub view.
& The rule requires ``default and key variants'' but never defines what a key variant is. The agent satisfied the literal requirement with one story and stopped. States like locked or already-applied themes — the most likely sources of user confusion — were never documented or accessibility-checked. \\
\hline
\end{tabular}
\caption{Vague Rule Examples}
\label{tab:vague_rules}
\end{table}

\subsection{Rules Followed for Coding Rules and Process Rules}
\label{app:techeval_codingandprocessrules}
\begin{table}[h]
  \centering
  \small
  \setlength{\tabcolsep}{4pt}
  \label{tab:freq}
  \begin{tabular}{ccl}
    \toprule
    Condition & \makecell{Process Rules \\ ($\mu$, 95\% conf)} & \makecell{Coding Rules \\ ($\mu$, 95\% conf)} \\
    \midrule
    No \appname & 0.35±0.20 & 0.59±0.10 \\
    Basic \appname & 0.49±0.14 & 0.59±0.13 \\
    Partial \appname\ & 0.72±0.20 & 0.66±0.08 \\
    Full \appname\ & \textbf{0.87±0.11*} & \textbf{0.75±0.07} \\
    \bottomrule
  \end{tabular}
\end{table}

\section{User Study Participants and Tasks}
\label{app:userstudy_participants}

\begin{table}[H]
\centering
\small
\setlength{\tabcolsep}{4pt}
\begin{tabular}{|c|p{2.5cm}|p{4cm}|p{4cm}|p{4cm}|}
\hline
\multicolumn{1}{|c|}{\textbf{Participant}} & 
\multicolumn{1}{c|}{\textbf{Age, Gender, Occupation}} & 
\multicolumn{1}{c|}{\textbf{Repository and Rules}} & 
\multicolumn{1}{c|}{\textbf{Task 1}} & 
\multicolumn{1}{c|}{\textbf{Task 2}} \\
\hline
P1 & 23F, SWE @ Big Tech & \textsc{Poneglyph} & Summarize Journal Entries & Auto-organize journal entries \\
\hline
P2 & 29F, SWE @ Big Tech & Full-stack passion project for teaching ML to HS kids & Create a new learning model & Use rules to clean up repo \\
\hline
P3 & 27F, Data Analyst @ Startup & Sandboxed repo of consumer data and analysis & Create plots and a business report from existing SQL queries & Run another SQL $\rightarrow$ Insights pipeline \\
\hline
P4 & 27F, SWE @ Big Tech & \textsc{Poneglyph} & Select a prompt and create dynamic responses & Auto-tag journal entries \\
\hline
P5 & 27M, 5th year PhD in ML & ML research project repo & Translate JAX to PyTorch & Modularize translation components \\
\hline
P6 & 26M, 4th year PhD in Computer Graphics & Graphics research project: heatmaps for contrast/luminance & Display-quality heatmap of contrast vs peak luminance & Scatterplot of display parameters on heatmap \\
\hline
P7 & 26M, 2nd year PhD in Econ & Econ research project on lightning strike data & Correlational analysis of strike time vs current & Visualize spatial correlation across BC with heatmap \\
\hline
P8 & 28M, SWE @ Startup & iOS text-to-speech project & Allow AI chat history to persist & Chunk paper text with visual signals/icons \\
\hline
P9 & 25F, SWE @ Big Tech & \textsc{Poneglyph} & Create settings component (language/aesthetics) & Allow image uploading \\
\hline
P10 & 26F, SWE @ Startup & \textsc{Poneglyph} & GPT-5.4 generate image/comic strip from journal & Create landing page \\
\hline
P11 & 25M, SWE @ Startup & \textsc{Poneglyph} & Rich text editing (MS Word-like) & Rich image editing \\
\hline
P12 & 27F, 3rd year PhD in ML & ML research project repo & Translate part of paper to code & Translate next part of paper to code \\
\hline
\end{tabular}
\caption{Participant demographics, repositories, and tasks.}
\label{tab:participants}
\end{table}









\end{document}